\documentclass{ws-ijmpa-FRK}

\def\beq{\begin{equation}}
\def\eeq{\end{equation}}
\def\ber{\begin{eqnarray}}
\def\eer{\end{eqnarray}}
\def\l{\Lambda}
\def\lsim{\
  \lower-1.5pt\vbox{\hbox{\rlap{$<$}\lower5.3pt\vbox{\hbox{$\sim$}}}}\ }
\def\gsim{\
  \lower-1.5pt\vbox{\hbox{\rlap{$>$}\lower5.3pt\vbox{\hbox{$\sim$}}}}\ }
\def\apj{{Astroph.\@ J.\ }}
\def\mn{{Mon.\@ Not.\@ Roy.\@ Ast.\@ Soc.\ }}
\def\asta{{Astron.\@ Astrophys.\ }}
\def\aj{{Astron.\@ J.\ }}
\def\prl{{Phys.\@ Rev.\@ Lett.\ }}
\def\prd{{Phys.\@ Rev.\@ D\ }}
\def\pd{{Phys.\@ Rev.\@ D\ }}

\def\plb {{Phys.\@ Lett.\@ B\ }}
\def \jetpl {JETP Lett.\ }
\def\etal{{\it et al.}}

\def\de {dark energy~}
\def\cc {{cosmological constant~}}
\def\atridot{\stackrel{...}{a}}

\def\omt{\Omega_{\rm m 0}}
\def\omx {\Omega_X}
\def\om{\Omega_{\rm m}}
\def\half{{1\over 2}}
\def\statei{\lbrace r,s\rbrace}

\def \lleq {\lower0.9ex\hbox{ $\buildrel < \over \sim$} ~}
\def \ggeq {\lower0.9ex\hbox{ $\buildrel > \over \sim$} ~}
\def\l{\Lambda}

\def\lsim{\
  \lower-1.5pt\vbox{\hbox{\rlap{$<$}\lower5.3pt\vbox{\hbox{$\sim$}}}}\ }
  \def\gsim{\
    \lower-1.5pt\vbox{\hbox{\rlap{$>$}\lower5.3pt\vbox{\hbox{$\sim$}}}}\ }

\begin{document}

\markboth{V. Sahni and A. Starobinsky}
{Reconstructing Dark Energy}

\catchline{}{}{}{}{}

\title{Reconstructing Dark Energy}

\author{Varun Sahni$^a$ and Alexei Starobinsky$^b$}
\address{$^a$Inter-University Centre for Astronomy and Astrophysics, 
Post Bag 4, Ganeshkhind, Pune 411~007, India \\varun@iucaa.ernet.in\\
$^b$Landau Institute for Theoretical Physics, Kosygina 2,
Moscow 119334, Russia\\alstar@landau.ac.ru}

\maketitle

\bigskip

\begin{abstract}
This review summarizes recent attempts to reconstruct the expansion history of
the Universe and to probe the nature of dark energy. Reconstruction methods 
can be broadly classified into parametric and non-parametric approaches. 
It is encouraging that, even with the 
limited observational data currently available, different approaches give 
consistent results for the reconstruction of the Hubble parameter $H(z)$ and 
the effective equation of state $w(z)$ of dark energy. Model 
independent reconstruction using current data allows for modest evolution of 
dark energy density with redshift. However, a cosmological constant (= dark 
energy with a constant energy density) remains an excellent fit to the data. 
Some pitfalls to be guarded against during cosmological reconstruction are 
summarized and future directions for the model independent reconstruction of dark 
energy are explored.
\end{abstract}

\section{Introduction}

The accelerated expansion of the universe has now been confirmed by several
independent observations including those of high redshift type Ia supernovae, 
and the cosmic microwave background (CMB) combined with the large scale 
structure of the Universe \cite{supernovae,gold,snls,wmap,tegmark}. Another
way of presenting this kinematic property of the Universe is to postulate 
the existence of a new entity -- dark energy (DE). This 
latter statement is dynamical in nature and therefore requires some
assumptions to be made about the form of gravitational field equations governing the 
evolution of the (observed part of the) Universe.
 
Although observationally well established, no single theoretical model provides
an entirely compelling framework within which cosmic acceleration or DE 
can be understood. Indeed, the very many models of DE existing in the 
literature illustrate that its nature is still very much an enigma. At present, 
all existing observational data are in agreement with the simplest possibility 
of DE being a cosmological constant $\Lambda$ with $\rho_{\Lambda}= 
\Lambda/8\pi G = const \simeq 10^{-47}$GeV$^4$ (inside $\sim 2\sigma$ error 
bars in the worst case).\footnote{$\hbar = c =1$ is used throughout the paper.}
This case is internally self-consistent and non-contradictory. The extreme
smallness of the cosmological constant expressed in either Planck, or even 
atomic units means only that its origin is {\em not} related to strong, 
electromagnetic and weak interactions (in particular, to the problem of the 
energy density of their vacuum fluctuations).\footnote{However, the
empirical relation $\rho_{\Lambda}\sim m_{\nu}^4$ where $m_{\nu}$ is some
characteristic neutrino rest-mass (the lightest one ?) suggests that vacuum 
energy of the interaction responsible for non-zero neutrino rest-masses may be 
relevant for a non-zero $\Lambda$.} Although in this case DE reduces to only a single 
fundamental constant we still have no derivation from any underlying 
quantum field theory for its small value. 

Within this context it is but natural that
other possibilities admitting a (slightly) variable dark energy 
have also been actively studied by the 
scientific community in recent years. Moreover, it is interesting that
properties of the currently observed 
`late DE' are {\em qualitatively} similar to those of an `early DE' which 
is believed to have given rise to accelerated expansion (inflation)
in the early Universe. However in the case of the latter, there are sufficient grounds
to support the view that
`early DE' was unstable and, thus, more complicated than
a cosmological constant. So, it is natural to conjecture by analogy that 
the same might also be true of `late DE'.

DE models proposed to account for the present cosmic 
acceleration include:\\ 
(i) Quiessence with $w \equiv p_{DE}/\rho_{DE}= {\rm constant}$,
the \cc $\Lambda$ ($w = -1$) is a special member of this class.\\
(ii) Quintessence models which are inspired by the simplest class of 
inflationary models of the early Universe and employ a scalar field rolling 
down a potential $V(\phi)$ to achieve late-time acceleration. Quintessence 
potentials with $V^{\prime   \prime} V/(V^{\prime})^2 \geq 1$
have the attractive property that \de approaches a common evolutionary
`tracker path' from a wide range of initial conditions.\\
(iii) The Chaplygin gas model (CG) has the equation of state 
$p \propto -1/\rho$ and evolves as $\rho=\sqrt{ A+B (1+z)^6}$ where $z$ is
the redshift, $z\equiv a(t_0)/a(t) - 1$. It therefore behaves like dark matter 
at early times ($z\gg 1$) and like the \cc at late times. 
CG appears to be the simplest model attempting to unify DE and non-baryonic cold dark
matter.\\
(iv) `Phantom' DE ($w < -1$). \\
(v) Oscillating DE. \\
(vi) Models with interactions between DE and dark matter.\\ 
(vii) Scalar-tensor DE models.\\
(viii) Modified gravity DE models in which the gravitational Lagrangian is 
changed from $R$ to $F(R)$ where $R$ is the scalar curvature and $F$ is an
arbitrary function. \\
(ix)  Dark energy driven by quantum effects. \\
(x) Higher dimensional `braneworld' models in which acceleration is caused
by the leakage of gravity into extra dimensions. \\
(xi) Holographic dark energy, etc.\\
See the reviews \cite{ss00,DE_review} for an exhaustive list of models and 
references. However, none of these models leads to the reduction of the
number of fundamental constants (parameters in the microscopic Lagrangian)
as compared to standard $\Lambda$CDM. In other words, at the current
state-of-the-art, DE requires at least one new parameter 
whose value is set from observations.

Models with variable DE can be broadly divided into two main classes: 

\smallskip
{\tt 1.} {\underline{\em Physical DE}}, in these models DE is the energy density of some new, very
weakly interacting physical field. 

{\tt 2.} {\underline{\em Geometrical DE }} (otherwise dubbed modified gravity models). 
In these models the
gravity equations do not coincide with those of Einsteinian general 
relativity. However, it is usually possible to re-write the new equations in the conventional
{\em Einsteinian} form by transferring all additional terms
from the l.h.s. 
into the r.h.s. of the Einstein equations and referring to them as an
effective energy-momentum
tensor of DE (see Sec. 2 below). (Precisely this happened to the cosmological constant
which originally appeared in the l.h.s. of Einstein's field equations but is now felt
by many to constitute an effective matter term such as vacuum energy, etc.)

Another category, which is even more important from the observational
point of view, arises in response to the question whether or not 
the description of DE requires a new field degree of
freedom (= a new kind of matter). If the answer is in the affirmative then DE may be
considered as being `induced' by other kinds of matter. All physical DE models and
many geometrical ones belong to this category but there do exist geometrical DE
models which do not (for instance the $F(R)$ model with the Palatini
variation of its action).

Faced with the increasing proliferation of DE models each with its own physical
motivations and assumptions, a concerned cosmologist can proceed in either of 
two ways: 

(i) Test {\em every single} model against observations.

(ii) Try and ascertain properties of dark energy in a {\em model independent 
manner}.

In this article, we proceed along route (ii) and attempt to review both the
successes as well as difficulties faced by methods attempting to reconstruct
the properties of dark energy directly from observations in a model 
independent manner.

\section{Model independent reconstruction of Dark Energy}

Before attempting to determine its properties, we first need to provide a 
definition 
of {\em dark energy}. 
A traditional approach is to use the Einstein form 
of the gravitational field equations
\beq\label{Einst}
R_{\mu\nu} - {1\over 2}g_{\mu\nu}R = 8\pi G \left(\sum_a T_{\mu\nu}^{(a)}
+ T_{\mu\nu}^{DE}\right)
\eeq
as providing a {\em definition} of the effective energy-momentum tensor
$T_{\mu\nu}^{DE}$ of DE. Here, the summation over $a$ in the r.h.s. includes all
types of matter known from laboratory experiments (protons, neutrons, photons,
neutrinos, etc.) as well as non-relativistic non-baryonic cold dark matter
(whose energy-momentum tensor is dust-like in the first approximation
$0<p \ll \rho$). $G=G_0=const$ is the present value of Newton's 
gravitational constant.

This definition has a number of advantages: \\
(i) it is simple, well-defined and self-consistent; \\
(ii) it treats physical and geometrical DE on an equal footing; \\
(iii) in the absence of direct physical interaction between DE with 
known forms of matter or with cold dark matter, the DE energy-momentum tensor
is conserved: $T_{\mu ;\nu}^{\nu (DE)}=0$. \\
One should stress here that
using (\ref{Einst}) we automatically ascribe terms describing (possible) gravitational
interactions between DE and non-relativistic matter, as well as
the matter energy-momentum tensor multiplied by a change in the effective
gravitational constant, to the DE energy-momentum tensor. The latter
possibility arises, for instance, in scalar-tensor DE models.  
  
When applied to a homogeneous and isotropic Friedmann-Robertson-Walker (FRW)
cosmological model, eqs. (\ref{Einst}) reduce to two algebraically 
independent equations:
 \ber
H^2 &=& \frac{8\pi G}{3}\left(\sum_a \rho_a +\rho_{DE}\right) - 
\frac{k}{a^2}~,\label{eq:acc}\\
\frac{\ddot a}{a} &=& -\frac{4\pi G}{3}\left(\sum_a (\rho_a
+ 3p_a) + \rho_{DE} + 3p_{DE}\right) 
\label{eq:acc1}
\eer 
where $a(t)$ is the FRW scale factor and the Hubble parameter $H(t)\equiv 
\dot a/a$.

At late times when radiation may be neglected, one gets
\beq\label{eq:energy}
\rho_{\rm DE} =
\frac{3H^2}{8\pi G}(1 - \Omega_{\rm m})
\eeq
where we have omitted the contribution from the curvature term in 
(\ref{eq:acc}) for simplicity. $\Omega_m$ is the total density of
non-relativistic matter in terms of its critical value. Similarly, the 
expression for the deceleration parameter 
\beq
q\equiv -\ddot a/aH^2 = \frac{H'(x)}{H(x)}~ x - 1 \, , ~~~ x = 1 + z \, 
\label{eq:decel1}
\eeq 
(prime implies differentiation with respect to $x$) takes the 
form:
\beq
p_{\rm DE} = \frac{H^2}{4\pi G}(q - \frac{1}{2})~.
\label{eq:pressure}
\eeq

Dividing (\ref{eq:pressure}) by (\ref{eq:energy}) we get the following
expression for the {\em effective} equation of state (EOS) of dark energy ($w \equiv 
p_{DE}/\rho_{DE}$) :
\beq\label{eq:state}
w(x) = {2 q(x) - 1 \over 3 \left( 1 - \Omega_{\rm m}(x) \right)}
\equiv \frac{(2 x /3) \ d \ {\rm ln}H \ / \ dx - 1}{1 \ - \ (H_0/H)^2 
\Omega_{m0} \ x^3}\,\,.
\eeq
For physical DE, the EOS makes physical sense. However, this is not so for 
geometrical DE for which the acceleration of the Universe is caused by the fact 
that the field equations describing gravity are not Einsteinian.


As a specific example of geometrical DE consider the widely studied DGP braneworld \cite{DGP}
for which 
\beq\label{eq:dgp}
H = \sqrt{\frac{8\pi G \rho_{\rm m}}{3} + \frac{1}{l_c^2}} +
\frac{1}{l_c}~,
\eeq
where $l_c = m^2/M^3$ is a new length scale and $m$ and $M$ refer
respectively to the four and five dimensional Planck mass. The acceleration of 
the universe in this model arises not because of the presence of DE
but due to the fact that gravity becomes five 
dimensional on length scales $R > l_c = 2H_0^{-1}(1-\Omega_{\rm m})^{-1}$. 
The contrast between (\ref{eq:acc}) and (\ref{eq:dgp}) makes it abundantly
clear that for models such as DGP the EOS in (\ref{eq:state})
is an {\em effective} quantity ($w \equiv w_{\rm eff}$), which may
still be useful for descriptive purposes but which no longer represents any 
fundamental physical property of an accelerating universe. Indeed, instances 
are known when $w_{\rm eff} < -1$ even when matter itself satisfies the 
weak energy condition $\rho + P \geq 0$ \cite{beps00,ss02,gprs06}. It may be 
instructive to note that, for geometrical DE, $w(z)$ may show pathological 
behaviour in certain cases, such as the presence of poles at finite values of 
redshift, $w(z_p \to \pm \infty)$, even though the underlying cosmological 
model is completely well behaved (see for instance 
\cite{linder04,loiter}). For such models, the deceleration parameter $q(z)$ 
and other geometrical parameters prove to be more robust 
for determining DE properties than the EOS. We shall return to these 
important issues in Section \ref{sec:statefinder}.

Observational tests of DE rely on an accurate measurement of at least one of the 
following quantities:

\begin{enumerate}

\item The luminosity distance
\beq\label{eq:lumdis}
\frac{D_L(z)}{1+z} =  \int_0^{z} \frac{dz'}{H(z')}~.
\eeq

\item The angular size distance
\beq\label{eq:angdis}
D_A(z) =  \frac{1}{1+z}\int_0^{z} \frac{dz'}{H(z')}~.
\eeq

\item The coordinate distance
\beq\label{eq:coordis}
r(z) =  \int_0^{z} \frac{dz'}{H(z')}~.
\eeq

\end{enumerate}
Here, we have assumed that the universe is spatially flat for simplicity.
Note that the distance duality relation $D_L = (1+z)^2 D_A(z)$ which follows 
from (\ref{eq:lumdis}) and (\ref{eq:angdis}) is valid only for metric theories 
of gravity \cite{bassett_kunz}. Its violation (if observed) could, 
therefore, be used to probe alternative theories of gravity. Existing 
data, however, appear to support it (within observational errors) \cite{bgm06}.

In all of the above expressions the value of the Hubble parameter can be 
`reconstructed' through a relation similar to the one given below for the 
luminosity distance
\cite{st98,HT99,chiba99,saini00,chiba00}: 
\beq\label{eq:H}
H(z)=\left[{d\over dz}\left({D_L(z)\over 1+z}\right)\right]^{-1}~.
\eeq 
Differentiating a second time allows one to reconstruct the equation of state 
of DE (\ref{eq:state}). Equations (\ref{eq:state}) and (\ref{eq:H}) immediately
inform us that $w(z)$ will be a noisier quantity than $H(z)$ since two
successive differentiations are needed for the reconstruction $D_L \to w(z)$
while a single differentiation suffices for $D_L \to H(z)$. This has led 
several people to suggest $H(z)$ (or $\rho_{\rm DE}(z)$) as being better 
suited for providing a model independent description of properties of DE.
Another important difference between $H(z)$ defined in (\ref{eq:H}) and 
$w(z)$ in (\ref{eq:state}) is that the former is independent of the value
of the matter density parameter $\Omega_m$ while the latter is not. As a 
result, uncertainties in the current value of $\Omega_m$ affect the 
reconstruction of the EOS far more profoundly than they do $H(z)$.
We shall return to this issue in section \ref{sec:obstacle}.

Knowing $H(z)$ (either through (\ref{eq:H}) or using corresponding relations
for other observational tests), allows us to extend cosmological reconstruction
to other important physical properties of the Universe including:

\begin{itemize}

\item Its age 
\beq\label{eq:age}
t(z) = \int_z^\infty \frac{dz'}{(1+z') H(z')} \, .
\eeq

\item The deceleration parameter (\ref{eq:decel1}) and the equation of state
(\ref{eq:state});

\item the electron-scattering optical depth to a redshift $z_{\rm reion}$
\beq
\tau(z_{\rm reion}) = c\int_0^{z_{\rm reion}}\frac{n_e(z)\sigma_T
~dz}{(1+z)H(z)} \, , \label{eq:reion}
\eeq
where $n_e$ is the electron density and $\sigma_T$ is the Thomson cross-section
describing scattering between electrons and CMB photons.

\item The product $d_A(z)H(z)$, which plays a key role in the Alcock--Paczynski
anisotropy test \cite{alcock}.

\item The product $d_A^2(z)H^{-1}(z)$, which is used in the volume-redshift 
test \cite{davis}.

\item The parameter $A$ associated with the determination of the baryon 
acoustic peak \cite{eisenstein05}
\beq
A = \frac{\sqrt{\omt}}{h(z_1)^{1/3}}~\bigg\lbrack ~\frac{1}{z_1}~\int_0^{z_1}
\frac{dz}{h(z)}
~\bigg\rbrack^{2/3} = ~0.469 \pm 0.017~,
\eeq
where $h(z) = H(z)/H_0$ and $z_1 = 0.35$
is the redshift at which the acoustic scale has been measured in the redshift
sample.

\item The `shift' parameter $R$ associated with the CMB 
\cite{wmap,bond97,wangm06}

\beq
R = \sqrt{\omt}\int_0^{z_{\rm lss}}\frac{dz}{h(z)} = 1.7 \pm 0.03.
\eeq
\end{itemize}

 For quintessence, it is additionally possible to reconstruct its potential, 
since the Einstein equations 
{\setlength\arraycolsep{2pt}
\begin{eqnarray}
H^2 & = & \frac{8}{3}\pi G \left(\rho_m + {1\over 2}\dot\phi^2+ 
V(\phi)\right),\nonumber\\
\dot H &=& -4\pi\, G(\rho_m + \dot\phi^2)~,
\label{eqn:hsq}
\end{eqnarray}}
can be rewritten as (see, for instance, \cite{st98,saini00}):
{\setlength\arraycolsep{2pt}
\begin{eqnarray}
{8\pi G\over 3H_0^2} V(x)\ &=& {H^2\over H_0^2} 
-{x\over 6H_0^2}{dH^2\over dx} -{1\over 2}\omt\,x^3,
\label{eqn:Vzed}\\
{8\pi G\over 3H_0^2}\left({d\phi\over dx}\right)^2 &=& 
{2\over 3H_0^2 x}{d\ln H\over dx} 
-{\omt x\over H^2}, ~~ x\equiv 1+z~.
\label{eqn:phidot}
\end{eqnarray}}
Integrating (\ref{eqn:phidot}), one determines $\phi(z)$ to within an additive 
constant. The inversion $\phi(z) \to z(\phi)$ followed by substitution into 
(\ref{eqn:Vzed}) allows us to reconstruct $V(\phi)$, since $H(z)$ and its 
first derivative can be determined from observations using (\ref{eq:H}).
The presence of $\omt$ in (\ref{eqn:Vzed}) and (\ref{eqn:phidot})
implies that the value of the matter density must be known rather precisely
for an accurate reconstruction of $V(\phi)$. 

In several important cases (\ref{eqn:Vzed}) and (\ref{eqn:phidot}) have led
to a closed form expression for $V(\phi)$, for instance:

\begin{itemize}

\item DE with a constant equation of state $-1<w<0$ is described by
\cite{ss00,statefinder1}

\beq
V(\phi)={3H_0^2(1-w)(1-\Omega_{m0})^{1/|w|} \over 16\pi G\Omega_{m0}^\alpha}
\sinh^{-2\alpha}\left(|w|\sqrt{{6\pi G\over 1+w}}
(\phi-\phi_0+\phi_1)\right)~, 
\label{quequi}
\eeq
where
\beq
\alpha = \frac{1+w}{|w|},~~\phi_0=\phi(t_0),~~\phi_1= \sqrt{{1+w\over
6\pi G}}{1\over |w|} \ln {1+\sqrt{1-\Omega_{m0}}\over
\sqrt{\Omega_{m0}}}~. \nonumber
\eeq
Consequently, a universe consisting of such a scalar field will have
expansion properties which closely mimic
a different kind of DE, e.g., 
a tangled network of cosmic strings
($w=-1/3$) or domain walls ($w=-2/3$).

\item The Chaplygin gas model which unifies dark matter and DE and
has $p = - A/\rho$ can be described by the scalar field potential 
\cite{chap,chap1} (in the absence of additional cold dark matter and 
neglecting baryons and photons):

\beq
V(\phi) = \frac{\sqrt{A}}{2}\left (\cosh(2\sqrt{6\pi G}\phi) + \frac{1}
{\cosh(2\sqrt{6\pi G}\phi)}\right )~.
\eeq
Note that the behaviour of the
Chaplygin gas may also be modelled completely differently using a scalar field
with the Born-Infeld kinetic term \cite{bilic,frolov} (a specific type of k-essence). 
This illustrates once more that, even for physical DE, the equation of state $w(z)$ does not
uniquely define an underlying field-theoretical model
\footnote{The Born-Infeld Lagrangian ${\cal L} = -V_0\sqrt{1-\phi_{,\mu}\phi^{,\mu}}$
is quite different from the Quintessence Lagrangian ${\cal L} = \frac{1}{2}{\dot \phi}^2
- V(\phi)$.} 
(see also \cite{tirth03}).

\end{itemize}

From the fact that the left-hand side of Eq. (\ref{eqn:phidot}) is always 
non-negative, follows an important restriction on the expansion law for the 
Universe which must be satisfied before attempts to reconstruct the potential 
are made, namely
\beq
{dH^2\over dz}\ge 3\omt H_0^2(1+z)^2~.
\label{ineq}
\eeq
Equation (\ref{ineq}) is simply  a restatement of the weak energy
condition $\rho_{\phi}+p_{\phi}\geq 0$. Integrating (\ref{ineq}), we get the
relation
\beq
H^2(z)\ge H_0^2\left(1+\omt (1+z)^3\right)
\label{int-ineq}
\eeq
which is easier to verify observationally.

In the case of physical DE, with its implied minimal coupling to gravity, the expansion history 
$H(z)$ can also be reconstructed from 
the growth rate of inhomogeneous density perturbations in 
the non-relativistic matter component on scales significantly less than the Hubble radius
$H^{-1}$, (provided perturbations in the DE component can be neglected and the 
effective gravitational constant does not change with time) \cite{st98,ss00}. 
Indeed, the linearized perturbation equation \footnote{Galaxy peculiar velocities produced by these matter inhomogeneities
result in small angular dependent corrections to the luminosity distance
$D_L(z)$ and other quantities referring to the evolution of a FRW background that, in turn, can
be used to determine $H(z)$ \cite{BDG,BDK,EM}.}
\beq
{\ddot \delta} + 2H{\dot\delta} - 4\pi G\rho_m\delta = 0
\label{eq:recon3}
\eeq
can easily be inverted, with the result:
\beq
H(z)= H(0) \bigg\lbrack\frac{(1 + z)^2\delta'^2(0)}{\delta'^2(z)} -
3\omt\frac{(1 + z)^2}{\delta'^2(z)}\int_0^z
\frac{\delta\vert\delta'\vert}{1 + z}dz\bigg\rbrack^\half.
\label{eq:recon7}
\eeq
Furthermore, an interesting relationship follows between the current value
of the matter density $\omt$ and $\delta(z)$:
\beq
\omt = \delta'^2(0)\bigg(3\int_0^\infty\frac{\delta\vert\delta'\vert}{1
+
z}dz\bigg)^{-1},
\label{eq:recon6}
\eeq
which could provide a consistency check on direct observational determinations 
of $\omt$. At sufficiently low redshifts $z < 1$, the advent of deep redshift 
surveys probing large scale structure may help in reconstructing $H(z)$ using 
(\ref{eq:recon7}). Although it is unlikely that the value of $\delta(z)$ for 
$z > 1$ will be reliably known in the near future, the fact that the Universe 
is expected to become matter dominated at fairly low redshifts, $\om \to 1$ at 
$z \gg 1$, allows us to use the spatially flat matter dominated solution 
$\delta \propto (1 + z)^{-1}$ to extrapolate to higher redshifts and thereby
evaluate (\ref{eq:recon6}). 

As a result, for physical DE we now have two independent methods to reconstruct
the same quantity $H(z)/H_0$: (i) through quantities like $D_L(z)$ referring to
an unperturbed FRW background, and (ii) through $\delta(z)$ describing the growth
of perturbations on small scales. It therefore follows that one should be able to 
reconstruct the density perturbation $\delta(z)$ from observations of the 
luminosity distance.
Consider the quantity $E(z)=H_0D_L(z)/(1+z)$ which can also be written as
the difference of conformal times: $E(z)= a_0H_0(\eta(0)-\eta(z)), 
~\eta = \int dt/a(t)$.
For our purpose it will be useful to invert the quantity
$E(z)$ (determined from observations) to $z(E)$. Next, we again assume that 
DE does not cluster and rewrite (\ref{eq:recon3}) in the following integral 
form:
\beq
\delta(E(z)) = \delta(0) + \delta'(0)\int_0^E (1+z(E))dE
+\frac{3\Omega_m}{2}\int_0^E (1+z(E_1))dE_1\int_0^{E_1} \delta(E_2)dE_2 .
\label{eq:recondelta}
\eeq
Here $\delta' = d\delta/dE \equiv d\delta/dz$ since $E(z)\approx z$ for 
$z\ll 1$. Clearly (\ref{eq:recondelta}) can be solved iteratively if
the value of $\delta'(0)$ is known. The value of $\delta(0)$ does not really 
matter since we are only interested in the ratio $\delta(z)/\delta(0)$.
The requirement $\delta(z=\infty)=0$ leads to the integral condition:
\beq
\delta'(0)= - (3/2)\Omega_m\int_0^{E_{\rm lss}}\delta(E)dE~,
\eeq
and, as in the case of (\ref{eq:recon6}), one needs to postulate some 
reasonable interpolating behaviour for $\delta(E(z))$ at large redshifts for 
which no supernova data exist at present. Since $E_{\rm lss} = 
\int_0^{z_{\rm lss}} dz/h(z)$, the value of this quantity
should be possible to determine from 
the CMB.

Both (\ref{eq:recon7}) and (\ref{eq:recondelta}) can be used as consistency 
checks for physical DE which, by assumption, is minimally coupled to gravity. 
For geometrical DE on the other hand,
the linearized perturbation equation is usually modified from its conventional
form (\ref{eq:recon3}), see for instance \cite{beps00} for scalar-tensor gravity or
\cite{brane} for the DGP model. \footnote{The fact that in this case $\delta(z)$ is
expected to be
different from the reconstructed expression (\ref{eq:recondelta}) provides an interesting possibility
to distinguish between physical and geometrical models of DE.}
As a result, DE reconstruction becomes more
complicated. In particular, a generic scalar-tensor model of DE depends on two
arbitrary functions $V(\phi)$ and $F(\phi)$ -- the scalar field potential 
and its coupling to gravity respectively. In this case one needs information
from {\em both}
$D_L(z)$ and $\delta(z)$ in order to reconstruct the DE model unambiguously \cite{beps00}.

To summarize, we have shown in this section that the reconstruction of 
the expansion rate $H(z)$ and other DE properties from
observational data is well defined and unambiguous from the mathematical point
of view. In practice, however, the situation is much more difficult since
all observational functions such as $D_L(z)$ and $\delta(z)$ are noisy and
known only at discrete values of the redshift $\lbrace z_1,z_2 .... z_N
\rbrace$ (associated with the redshifts of $N$ supernovae in the 
case of $D_L(z)$). Thus, it is impossible to directly differentiate them with
respect to redshift as formulae (\ref{eq:H}) and (\ref{eq:recon7}) require.
Add to this the fact that the dispersion in the luminosity distance is not 
expected to get significantly better than \cite{snap04} $\sigma_{\ln d_L}=
0.07$, and one is confronted with a noisy quantity $D_L(z_i)$ sampled at a set 
of discrete intervals $\lbrace z_i\rbrace$. Therefore, to convert from 
$D_L(z_i)$ to the function $H(z)$ defined at {\em all} redshift values within 
the interval (say) $0 \leq z < 2$ using (\ref{eq:H}) clearly requires a crucial
additional step: some sort of {\em smoothing} procedure. This is usually 
accomplished using either parametric or non-parametric reconstruction which we turn to next.

\section{Parametric reconstruction}

This approach is based on the assumption that the quantities
$D_L(z), H(z), w(z)$ vary `sufficiently slowly' with redshift and can therefore
be approximated by a fitting formula (ansatz)
which relies on a small number of free parameters $a_i, ~i = 1, N$.
The ansatz for $D_L(z,a_i), H(z,a_i), w(z,a_i)$ is compared against
observations, and the values of the free parameters $a_i$ are determined using
a minimization procedure (usually maximum likelyhood).
Quite clearly a successful fitting function should, in principle,
be able to faithfully reproduce the properties of an entire class of DE models.
Different parameterizations have been suggested for:
$D_L$ \cite{st98,HT99,saini00,chiba00,repko06}, $H(z)$ \cite{statefinder1,statefinder2,alam04a,alam04b}, $w(z)$ \cite{polar,albrecht,efstathiou,maor02,copeland,linder,wangm,saini04,leandros,gong,Lazkoz,jassal05,feng06,uzan06}
and $V(z)$ \cite{simon,ohta}. 

We attempt to summarize some of these approaches below.

\begin{enumerate}

\item{\bf Fitting functions to the luminosity distance $D_L$ }

An approximation to a function can easily be generated by expanding it
in a Taylor series about some redshift $z_0$. 
Applied to the luminosity distance (with $z_0=0$),
this method gives the following
fitting function \cite{HT99}
\beq
\frac{D_L(z)}{1+z} = \sum_{i=1}^N a_iz^i\,\,.
\label{eq:poly}
\eeq 
Unfortunately, it appears that 
in order to accurately
determine quantities of interest such as $H(z)$ and $w(z)$ 
one must take a large number of terms in (\ref{eq:poly})
which greatly increases the errors of reconstruction \cite{albrecht}
and reduces the reliability of this ansatz.

A more versatile Pad{\`e}-type ansatz was suggested in \cite{saini00}
\beq
{H_0D_L(z)\over 1+z}
= 2\left[ \frac{x - A_1\sqrt{x} -1 + A_1}{A_2 x+
A_3\sqrt{x} + 2 - A_1 -A_2 -A_3}\right] ~, ~~ x = 1+z~,
\label{eqn:star}
\eeq
which is able to {\em exactly} reproduce the results both for CDM
($\Omega_m = 1$) and the steady state model ($\Omega_\l=1$).

Another accurate fit with a greater number of free parameters is \cite{chiba00}
\ber
&&\frac{H_0 D_L(z)}{1+z} = \eta(1)-\eta(y)~,\\
&&\eta(y)=2\alpha\left[y^{-8}+\beta y^{-6}+\gamma
  y^{-4}+\delta y^{-2}+\sigma\right]^{-1/8}~, ~~ y = 1/\sqrt{1+z}~.
\label{eq:fit_chiba}
\eer

\item{\bf Fitting functions to the DE density}

The dark energy density can be written 
as a truncated Taylor series polynomial in $x=1+z$,
$\rho_{\rm DE} = A_1 + A_2x + A_3 x^2$.
This leads to the following ansatz for the Hubble parameter \cite{statefinder1}
\beq\label{eq:poly_fit}
H(x) =  H_0\left\lbrack \om x^3 + A_1 + A_2x + A_3 x^2\right\rbrack^\half\,\, ,
\label{eq:taylor}
\eeq
which, when substituted in the expression for the luminosity distance
(\ref{eq:lumdis}), yields
\beq
\frac{D_L}{1+z} =  \frac{c}{H_0}\int_1^{1+z} \frac{dx}{\sqrt{\om x^3 +
A_1 + A_2 x + A_3 x^2}}\,\,.
\label{eq:taylor1}
 \eeq
This fitting function gives exact results for the cosmological 
constant $w = -1$ ($A_2 = A_3 = 0$) as
well as for quiessence with $w = -2/3$ ($A_1 = A_3 = 0$) and $w =
-1/3$ ($A_1 = A_2 = 0$).  The presence of the term $\om
x^3$ in (\ref{eq:poly_fit}) ensures that the ansatz correctly
reproduces the matter dominated epoch at early times ($z \gg 1$).
For quintessence models as well as the Chaplygin gas, the luminosity distance $D_L(z)$ 
in (\ref{eq:taylor1}) can be determined to an accuracy
of better than $1\%$ if $\omt \geq 0.2$ \cite{statefinder2}.

\item{\bf Fitting functions to the equation of state}

In this approach one assumes that the DE equation of state $w(z)$
is an unknown
variable whose behaviour is `guessed' by means of a suitable fitting 
function $w(z, a_i)$. Since 
\ber
H^2(z) &=& H_0^2 \lbrack \omt (1+z)^3 + \omx\rbrack^2~,\nonumber\\
\nonumber\\
\omx &=& (1-\omt) \exp{\left\lbrace 3 \int_0^{x-1} 
\frac{1+w(z,a_i)}{1+z} dz\right\rbrace }~,
\label{eq:hubble_recon}
\eer
the values of $a_i$ can be determined by comparing
the luminosity distance (\ref{eq:lumdis})
or $D_A(z), r(z)$, against observations.

Several possible fits for $w(z)$ have been suggested in the literature.
Perhaps the simplest is the Taylor expansion
\cite{albrecht}
\beq
w(z) = \sum_{i=0}^N w_iz^i\,\,,
\label{eq:wpol}
\eeq
which, for $N=1$, gives results significantly better than the Taylor
expansion for the luminosity distance (\ref{eq:poly}).
The simple two parameter representation $w(z) = w_0 + w_1z$ is however
of limited utility
since it is only valid for $z \ll 1$.
A considerably more versatile four parameter ansatz has been suggested in 
\cite{copeland}.

A popular two parameter fit was suggested by Chevallier and Polarski 
\cite{polar} and by Linder \cite{linder} 
\beq
w(a) = w_0+w_1(1-a) = w_0 + w_1\frac{z}{1+z}~,
\label{eq:cpl}
\eeq
the associated luminosity distance can be obtained by substitution into
(\ref{eq:hubble_recon}) and (\ref{eq:lumdis}).
A more general form for this fit is
\beq
w(a) = w_p + (a_p - a)w_a~,
\eeq
where $a_p$ is the value of the scale factor at the
`sweet spot' where the equation of state $w(a)$ is most tightly constrained.
The value of $a_p$ usually depends upon the data set being
used \cite{hu04,albrecht06a,linder06,albrecht06b}.

\end{enumerate}

\begin{figure*}
\begin{center}
$\begin{array}{c@{\hspace{0.15in}}c}
\multicolumn{1}{l}{\mbox{}} &
\multicolumn{1}{l}{\mbox{}} \\ [-0.2cm]
\epsfxsize=2.7in
\epsffile{h_exp_w_g.epsi} &
\epsfxsize=3.3in
\epsffile{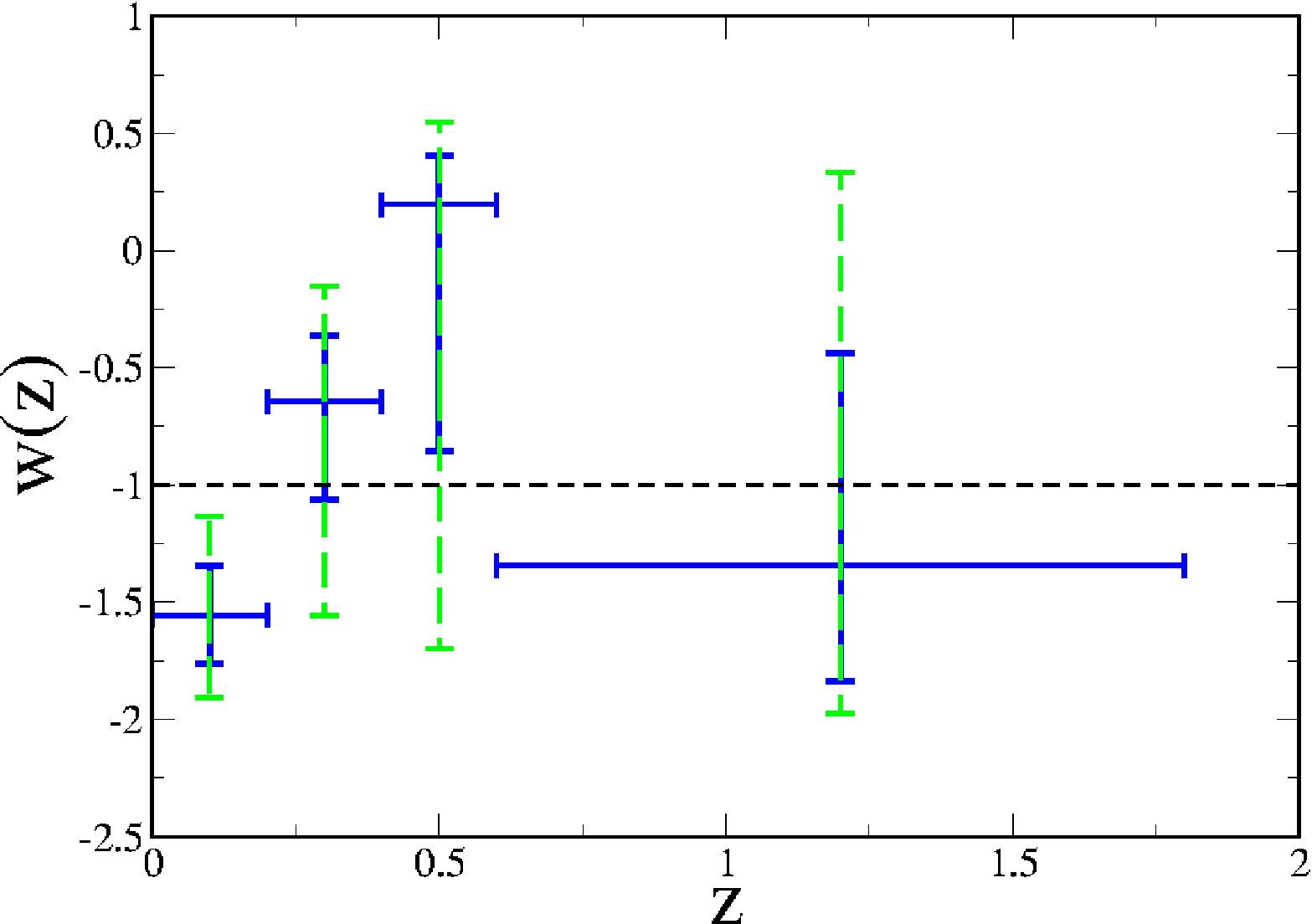} \\ 
\end{array}$
\end{center}
\caption{\small An early reconstruction of the supernova data (left panel)
using the parametric fit (\ref{eq:taylor}) shows an evolving equation of state 
to be marginally preferred over the cosmological constant
~\protect\cite{alam04b}. The same data were independently analyzed by means of 
a non-parametric ansatz (right panel) ~\protect\cite{huterer05}.
It is encouraging that both reconstructions appear to give similar results
for $0 < z \lleq 1$ where most of the data points lie. The crossing of the 
so-called `phantom divide' at $w=-1$ led to much theoretical interest and 
considerable model building activity. More recent SNe results, together with 
constraints from other measurements such as the CMB, LSS and Baryon acoustic 
oscillations imply less pronounced evolution of $w(z)$ with redshift as shown 
in figure \ref{fig:EOS}, see also ~\protect\cite{leandros,tirth05,gong,wangteg05,daly05,huterer05,paddy06,zhao06,otto06,hao_wei}. The left panel is from Alam, Sahni 
and Starobinsky ~\protect\cite{alam04b} while the right panel is from Huterer 
and Coorey ~\protect\cite{huterer05}.} \label{fig:hubble}
\end{figure*}

As mentioned earlier, for an ansatz to be regarded as being successful
it should embrace within its fold the behaviour of a reasonably wide
class of DE models. Most of the ansatz's discussed above successfully 
accommodate DE whose equation of state evolves moderately with redshift. It is quite clear 
that these simple fits cannot be used to rule out (using observations)
models with rapidly evolving $w(z)$. One reason for this is that applying an 
ansatz such as (\ref{eq:taylor}) or (\ref{eq:cpl}) to SNe data
is equivalent to smoothing the evolution of the Universe over a
redshift interval $\Delta z \propto 1/N$ where $N$ is the number of free
parameters in the ansatz (the concrete form of an ansatz defines the form of
a smoothing redshift function). Clearly, an implicit
smoothing such as this will cause 
rapid transitions in dark energy to disappear not because of
disagreement with data but simply because of the manner in
which the data have been `massaged' \cite{alam04a,alam04b}.
(An example of the disastrous results of applying the prior $w = $ constant
to models with an evolving EOS is discussed in section \ref{sec:obstacle}.)
Models 
with a fast phase transition in dark energy ($|{dw\over dz}|\gg 1$ over a
narrow range of redshift $\delta z \ll 1$) have been discussed in
\cite{bassett,corasaniti}.
To accommodate such models the following ansatz was suggested
\cite{bassett,bck04}
\beq
w(z)=w_i+\frac{w_f-w_i}{1+{\rm exp}(\frac{z-z_t}{\Delta})}\,\,,
\label{eq:bass} 
\eeq
where $w_i$ is the initial equation of state at high redshifts, 
$z_t$ is a transition redshift at which the equation of state falls to
$w(z_t) = (w_i + w_f)/2$ and $\Delta$ describes the rate of change of $w(z)$.
Substitution into (\ref{eq:hubble_recon}) and
(\ref{eq:lumdis}) gives the luminosity distance.

Finally, models with oscillating DE have also been discussed in the literature
\cite{oscillating_DE,leandros,Lazkoz,feng06,hao_wei}. 
To accommodate an oscillating EOS the following ansatz 
has been suggested \cite{feng06}
\beq
w(\log{a}) = w_0 + w_1\cos{\lbrack A\log{a/a_c}}\rbrack
\eeq
where $w_0, w_1, A, a_c$ are free parameters whose values must be obtained
by fitting to observations. 
It is encouraging to note that, in spite of some ambiguity in the form
of the different fits, when applied to the same supernova data set
most of them give consistent results in the
range $0.1 \lleq z \lleq 1$ (where there is sufficient data)
\cite{alam04a,alam04b,leandros,gong,huterer05}.

 One should also note that although increasing the number of parameters 
usually increases the accuracy of reconstruction of the `best fit', this is 
often accompanied by severe degeneracies which limit the utility of introducing
a large number of free parameters. In addition, extra free parameters are quite
severely penalized by information criteria such the Akaike information 
criterion \cite{akaike74} and the Bayesian information criterion 
\cite{schwarz78}, see also \cite{saini04,liddle04,bck04}. However, there 
exists a subtlety when applying these information criteria to 
fitting functions of
$H(z), w(z)$ etc. as opposed to
concrete theoretical models of DE.
In the case of the former, one should keep in mind the
possibility that a fit with a larger number of parameters might in fact
be describing the behaviour of a fundamental DE model (not yet known) 
containing a smaller number of truly free parameters.

The reverse is also true, a primitive fit (hence not penalizable) may, if the information
criteria are applied, detract our attention from a more complicated but also more
fundamental explanation of a phenomenon. Let us illustrate this with an example.
The theory of gravitational instability informs us that the linear density contrast
$\delta$ and the peculiar velocity field ${\bf v}$ are related as \cite{peebles80,colesrev}
\beq
\delta = -\frac{1}{aHf}{\bf \nabla}\cdot{\bf v}
\label{eq:velocity}
\eeq
where the function $f \equiv d\log\delta/d\log a$ can be approximated as
$f \simeq \Omega_m^{0.6}$. An exact calculation however reveals $f$ to be a more complicated
function involving elliptical integrals, etc. Therefore
a naive application of information criteria to (\ref{eq:velocity})
would needlessly penalize the latter, which 
is correct 
but complicated, in favour of the simpler $f \simeq \Omega_m^{0.6}$.

Another popular approach the so-called Principal Component analysis 
is based on expanding $w(z)$ in terms of a basis of orthogonal functions.
The form of these functions depends upon the kind of data used and its
constraining capabilities \cite{hut03,PC,huterer05}.
We end this section by noting that the reconstruction approach has also been
extended to scalar-tensor gravity \cite{beps00}, string inspired cosmology
with the Gauss-Bonnet term coupled
to a scalar field \cite{ishwar,nojiri06} and dark energy 
with non-canonical kinetic energy terms \cite{melchiorri}.
Other applications of cosmological reconstruction may be found in
\cite{recon}.

\begin{figure*}
\centerline{ \psfig{figure=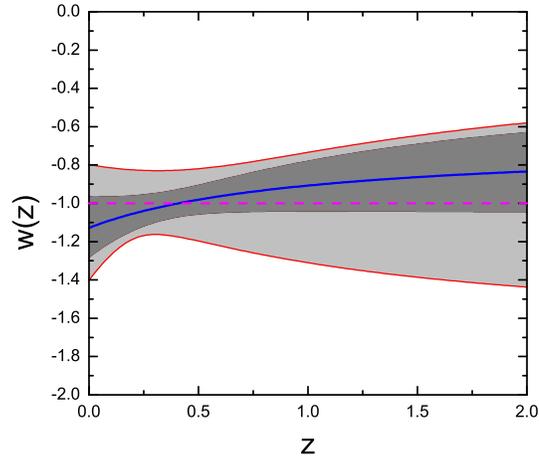,width=0.65\textwidth,angle=0} }
\caption{\small
The equation of state of dark energy $w(z)$ reconstructed using the WMAP 3 year
data +
157 ``gold'' SNIa data + SDSS. 
Median (central line), 68\%(inner, dark grey) and 95\%(outer, light
grey) intervals. The two parameter fit (\ref{eq:cpl}) has been used
in this exercise. From Zhao {\it et al.} ~\protect\cite{zhao06}.}
\label{fig:EOS} 
\end{figure*}

\section{Non-parametric reconstruction}

Non-parametric smoothing usually involves directly smoothing either $D_L$, or 
some other quantity appropriately binned in redshift space with some 
characteristic smoothing scale.
Different ways of implementing this approach have been discussed in
\cite{wang01,hut03,saini03,daly03,daly04,wangteg04,wangteg05,inverse,arman,daly05,huterer05,tavakol06}. All reconstruction  methods must deal with the 
fact that the data sample is usually sparse and the coverage of redshift space 
uneven. Consider for instance the comoving coordinate 
distance to a single supernova within a larger sample 
\beq
r_i = r(z_i) + n_i~.
\eeq
Here, $n_i$ is the noise term $\langle n_i\rangle = 0, \langle n_in_j\rangle = 
\sigma_i^2\delta_{ij}$, and $r(z)$ is related to the distance modulus $\mu_0$ 
of the supernova by \cite{wangteg05} $r(z)/{1{\rm Mpc}} = 
10^{\mu_0/5-5}/2997.9(1+z)$. Direct differentiation of $r_i$ to give 
$H^{-1}(z_i)$ will obviously result in a very noisy quantity. One way to tackle
this is to fit $r_i(z)$ piece-wise using a set of basis functions. Following 
this approach, Daly and Djorgovsky \cite{daly03,daly04} reconstructed the 
deceleration parameter from the dimensionless coordinate distance $y = H_0r$ 
by: (i) fitting $y(z)$ locally within a small redshift bin by means of 
a second order polynomial requiring that at least 10 data points lie within
each bin; (ii) determining the first and second derivatives of $y(z)$ from the
fit coefficients (for each bin) and reconstructing the deceleration parameter 
from $y(z)$ by means of the relation 
\beq
-q(z) \equiv \frac{{\ddot a}a}{{\dot a}^2} = 1 + (1+z)\frac{d^2y/dz^2}{dy/dz}~.
\eeq
Applying their method to a set of radio galaxies in addition to type Ia SNe, 
Daly and Djorgovsky \cite{daly03,daly04,daly06} found that although the cosmological 
constant provided a
very good fit to the data, modest evolution in the dark energy density was 
also perfectly acceptable. The transition from deceleration to acceleration 
occured at $z > 0.3$, with a best fit value of $z = 0.42$.
These results are in broad agreement with those obtained using parametric 
approaches
\cite{alam04a,alam04b,leandros,gong}.

A different approach to non-parametric reconstruction is discussed by
Shafieloo \etal \cite{arman} who
generalize a smoothing ansatz widely used in
the analysis of large scale structure. According to this method, a 
smoothed quantity
$D^S({\bf x})$ is constructed from a fluctuating `raw' quantity
$D({\bf x'})$ using a low pass filter $F$ having a smoothing scale
$\Delta$ 
\beq
D^S({\bf x}, \Delta) = \int D({\bf x'}) F(|{\bf x}-{\bf x'}|;
\Delta)~d{\bf x'}~.
\eeq
In large scale structure studies, $D$ is the density field 
\cite{coles,colesrev} whereas 
for cosmological reconstruction $D$ could be either of 
$D_L(z), D_A(z), r(z)$.
Commonly used filters include: (i) the `top-hat' filter, which has a
sharp cutoff $$F_{\rm TH} \propto \Theta\left (1-\frac{|{\bf x}-{\bf x'}|}
{\Delta}\right )~,$$
where $\Theta$ is the Heaviside step function ($\Theta(z) = 0$
for $z \leq 0$, $\Theta(z) = 1$ for $z > 0$) and (ii) the Gaussian
filter 
\beq  
  F_{\rm G} \propto \exp{\left (-\frac{|{\bf x}-{\bf x'}|^2}{2\Delta^2}
\right )}~.
\eeq
When applied to SNAP-quality data using a Gaussian filter, this method 
reconstructs the Hubble parameter to an accuracy of $\lleq 2 \%$
within the redshift interval $0<z<1$. 
The look-back time
\beq
T(z)=t(z=0)-t(z)=H_{0}^{-1}\int_{0}^{z} \frac{dz'}{(1+z') H(z')} 
\eeq
is reconstructed to an even better accuracy of $\lleq 0.2 \%$ at 
$z \simeq 1.7$.

\begin{figure*}
\centerline{ \psfig{figure=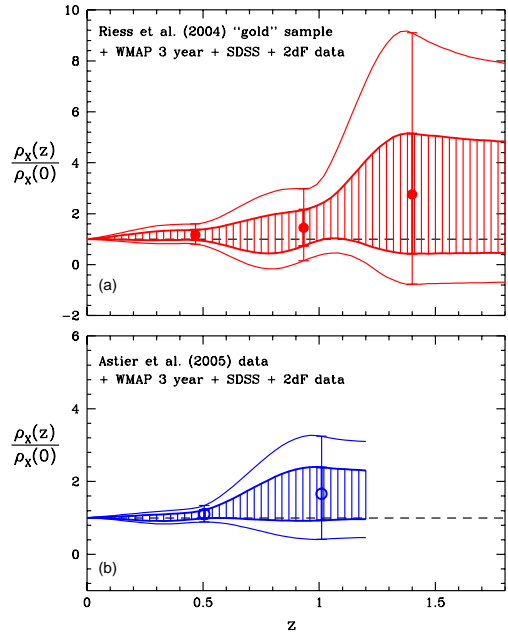,width=0.7\textwidth,angle=0} }
\caption{\small
Dark energy density $\rho_X(z)$ reconstructed using SN Ia data 
~\protect\cite{gold,snls},
combined with the WMAP 3 year data, the SDSS baryon acoustic oscillation data,
and the 2dF linear growth data for density perturbations.
The 68\% (shaded) and 95\% confidence contours are shown.
Beyond $z_{\rm cut}$=1.4 (upper panel) and 1.01 (lower panel),
$\rho_X(z)$ is parametrized by a power law $(1+z)^\alpha$.
The horizontal dashed line shows the unevolving density associated with a
cosmological constant. From Wang and Mukherjee ~\protect\cite{wangm06}.}
\label{fig:wangm}
\end{figure*}

Despite the considerable success of these approaches, neither they, nor the 
parametric methods discussed earlier provide independent measurements of 
either $H(z)$ or $q(z)$ within a given redshift range. One approach towards an 
uncorrelated determination of cosmological quantities is to bin the data in 
redshift bins of size $\Delta$ with a weight (or window) function assigned to 
each bin \cite{wangteg05}. Each window function vanishes outside its redshift 
bin and the two supernovae falling on either side of a bin boundary are 
discarded. This method ensures that, since two adjacent window functions have 
no common supernovae, the measurement of quantities within these bins will 
have uncorrelated error bars. This method is quite successful and complements 
other reconstruction approaches in many respects. One weakness, which 
practitioners of this approach have to guard against
is that, since some information at the bin boundary is lost,
discontinuities in the fit values at bin edges may be present \cite{daly03}.
These, in turn, could lead to unphysically large derivatives of derived 
quantities such as $H(z)$ and $w(z)$. 
Results obtained by applying a variant of
this method to recent CMB+SNe+LSS data are shown in
figure \ref{fig:wangm} and appear to
agree with the results of the parametric approach in figure \ref{fig:EOS}.
Other promising methods for obtaining uncorrelated estimates of
cosmological evolution are discussed in \cite{huterer05}.

In all of the above approaches, it is important to choose the value of the 
smoothing scale $\Delta$ (or bin size) optimally, so that 
(i) a sufficient number of data points is accommodated
within each bin to reduce the effects of shot noise, (ii) the bin size is not 
too large to cause excessive smoothing. The following useful formula 
giving the relative error
bars on $H(z)$ comes to our aid \cite{tegmark02}
\beq
\label{eq:tegmark}
\frac{\delta H}{H}\propto \frac{\sigma}{N^{1/2} \Delta^{3/2}}\,\,,
\eeq 
where N is the total number of supernovae (assuming an approximately uniform
distribution) and $\sigma$ is
the noise of the data. Clearly, a reduction in
the value of $\Delta$ by 3 should be compensated for by having
27 many times more supernovae if one wants to keep $\frac{\delta H}{H}$ 
unchanged. The situation is far worse for $w(z)$ which has a 
$\Delta^{-5/2}$ scaling, since the data are being differentiated twice to get 
$w(z)$.

\section{Obstacles to cosmological reconstruction}
\label{sec:obstacle}

\begin{figure*}
\centerline{ \psfig{figure=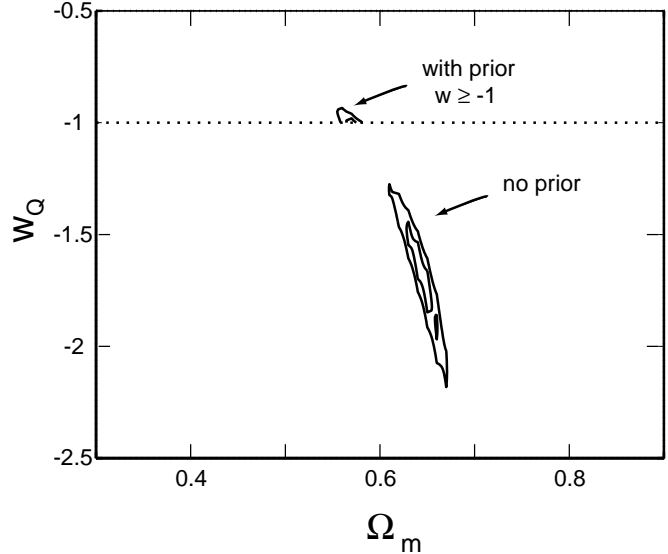,width=0.70\textwidth,angle=0} }
\bigskip
\caption{\small Pitfalls in cosmological reconstruction 1.
This figure from Maor \etal \protect\cite{maor02} shows the results of a 
reconstruction exercise performed by
assuming: (i) $w_{\rm Q}$ = constant as a prior, which gives
the larger lower contour with
$w_{\rm Q} < -1$; (ii) the additional constraint
$w_{\rm Q} \geq -1$, results in the smaller upper contour with
$w_{\rm Q} = -1$ as the best fit.
Both (i) and (ii) give confidence contours and best fit values of
$w_{\rm Q}$ and $\Omega_m$ which are widely off the mark since they differ 
from the fiducial Quintessence model which has 
$w_{\rm Q}(z) = -0.7 + 0.8 z$ and $\Omega_m = 0.3$.
}
\label{fig:maor}
\end{figure*}

With respect to cosmological reconstruction, considerable caution must be 
exercised when setting priors on the values of cosmological parameters.
This is true both for the density parameter $\Omega_m$ whose value is 
currently known to about $15\%$ accuracy and the DE equation of state.
Maor \etal \cite{maor02} gave a particularly insightful example of the dangers 
of incorrect
reconstruction by assuming a fiducial DE model with $\Omega_m = 0.3$
and an evolving EOS 
$w_{\rm Q}(z) = -0.7 + 0.8 z$. 
In their reconstruction, the results of which are shown in figure 
\ref{fig:maor}, it was assumed that $w_{\rm Q}$ was a constant and, 
additionally, that $w_{\rm Q} \geq -1$. These two (incorrect) priors led to a 
gross underestimation of $w_{\rm Q}$ and an overestimation of $\Omega_m$ as 
shown in figure \ref{fig:maor}.

\begin{figure*}
\centerline{ \psfig{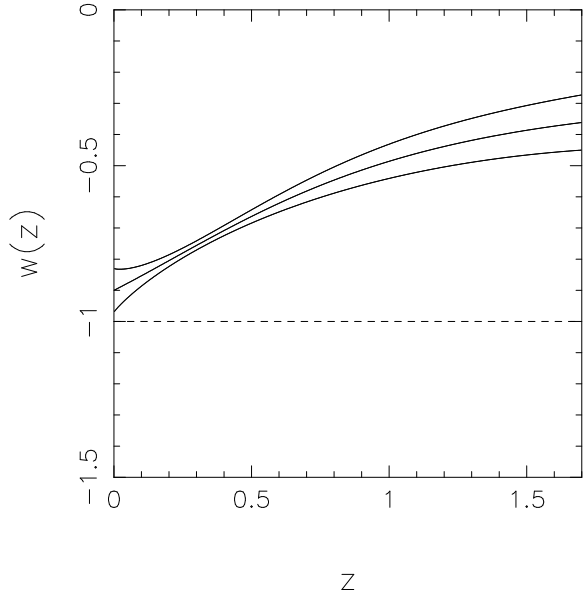} }
\bigskip
\caption{\small Pitfalls in cosmological reconstruction 2.
The reconstructed equation of state $w(z)$ 
is shown for $1000$ realizations of
an $\Omega_m=0.3, w=-1$, $\Lambda$CDM model assuming SNAP quality data. 
An {\em incorrect} value for the matter
density, $\omt=0.2$, is assumed in the reconstruction exercise
which uses the polynomial ansatz (\ref{eq:taylor}). 
The dashed line represents the fiducial
$\Lambda$CDM model with $w=-1$ while the solid lines 
show the mean value of the (incorrectly) 
reconstructed $w(z)$ and $1\sigma$ confidence levels around the mean.
Note that the reconstructed EOS excludes the fiducial $\Lambda$CDM model
to a high degree of confidence.
From Shafieloo \etal ~\protect\cite{arman}.
 } 
 \label{fig:arman}
 \end{figure*}

The above example shows that the EOS of dark energy can be badly reconstructed if we
assume a misleading prior for $w$. Since $w(z)$ depends upon the value of 
$\Omega_m$ through (\ref{eq:state}), it follows that an incorrect assumption 
about the value of the matter density has the potential to affect cosmological 
reconstruction quite significantly. This is demonstrated in figure 
\ref{fig:arman} in which the DE EOS is reconstructed
for a fiducial $\Lambda$CDM model with $w(z) = -1$ and $\Omega_m = 0.3$.
Cosmological reconstruction is based on the ansatz (\ref{eq:poly_fit}) in 
which the {\em incorrect} value $\Omega_m = 0.2$ is assumed.
This results in the reconstructed EOS evolving with redshift when, in fact,
no such evolution is present in the fiducial model !

\section{Looking beyond the equation of state}
\label{sec:statefinder}

\subsection{The $w${\tt -probe}}

As we have seen, the EOS has certain blemishes  -- it is obtained from $D_L$ 
after differentiating {\em twice}  and it is sensitive to the {\em precise} 
value of $\omt$. Both these features hamper its reconstruction which has led 
several authors to propose the Hubble parameter or the DE density as being 
more appropriate for cosmological reconstruction \cite{statefinder1,daly03,daly04,wangm,wangteg05}. Carrying these ideas further, the possibility of 
extracting information about the equation of state from the reconstructed 
Hubble parameter by constructing a {\em weighted average} of the equation of 
state was explored in \cite{alam04a}. This quantity, dubbed the 
$w${\tt -probe}, gleans information about the equation of state from the 
{\em first derivative} of the luminosity distance. Therefore, it is less noisy 
and better determined than $w(z)$.

The weighted average of the equation of state is defined as \cite{alam04a}
\beq
\label{eq:weight1}
1+\bar{w}= \frac {1}{\delta \ln(1+z)}\int \left(1+w(z)\right)
\frac{dz}{1+z}\,\, .
\eeq
An important feature of the $w${\tt -probe} ($\bar{w}$) is that it
can be expressed in terms of the difference in dark
energy density over a given redshift range
\ber\label{eq:weight2}
1+\bar{w}(z_1,z_2) &=& \frac{1}{3} \frac{\delta \ \ln \tilde{\rho}_{DE}}
{\delta \ \ln (1+z)}~~\nonumber\\
&\equiv& \frac{1}{3}{\rm ln}\left[\frac{H^2(z_1)-\Omega_{0\rm m} (1+z_1)^3}
{H^2(z_2)-\Omega_{0\rm m} (1+z_2)^3}\right] {\Big /} {\rm ln}
\left(\frac{1+z_1}{1+z_2}\right)~.\nonumber\\
\eer
Here $\delta$ denotes the total change of a variable between
integration limits and $\tilde {\rho}_{DE}=\rho_{DE}/\rho_{0c}$ 
($\rho_{0c}=3H^2_0/8\pi G$). Thus $\bar{w}$ is easy to determine if the Hubble
parameter has been accurately reconstructed.

\begin{figure*}
\centerline{ \psfig{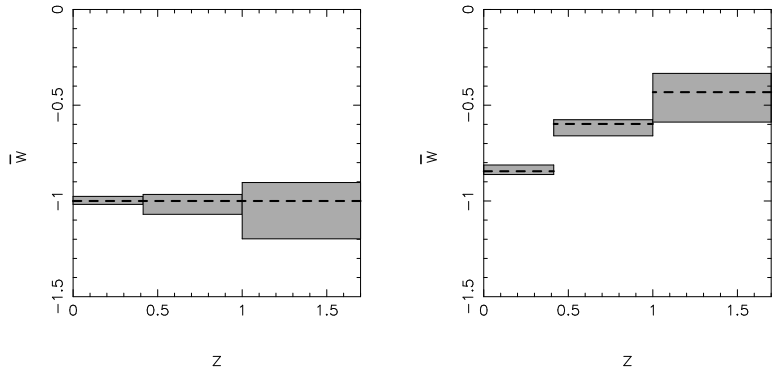} }
\caption{\small
The $w${\tt -probe} is reconstructed for the $\Lambda$CDM model
with $w=-1$ (left) and an evolving DE model with $w=-1/(1+z)$
(right). $1000$ realizations of SNAP-like data have been used.
The thick dashed line in both panels indicates the exact value of
$\bar{w}$ for the fiducial model, the dark grey boxes in each panel
indicate the $1\sigma$ confidence levels on $\bar{w}$ reconstructed
for the two models after
marginalising over $\omt=0.3 \pm 0.07$. The results of both panels show that
the $w${\tt -probe} is able to accurately determine DE properties even
if the matter density is not perfectly known.
From Shafieloo \etal ~\protect\cite{arman}.}
\label{fig:margin}
\end{figure*}

An important property of $\bar{w}$ is that it is less sensitive to 
uncertainties in the value of $\omt$ than $w(z)$.
This is demonstrated in figure \ref{fig:margin}
where the results for $\bar{w}$ are shown
after marginalizing over the matter density.
Remarkably, the value of $\bar{w}$ for the fiducial $\Lambda$CDM model remains
close to $-1$, while $\bar{w}$ for the evolving DE model
shows a clear signature of evolution. Thus, small uncertainties in the 
value of the matter density
do not appear to adversely affect
the accuracy of the reconstructed $w${\tt -probe}.
Furthermore, several excellent methods
for determining $\rho_{\rm DE}$ and $H(z)$ have been suggested in the
literature \cite{daly03,daly04,statefinder1,alam04a,wangm,wangteg05}, 
any of which could be used to determine $\bar{w}$ using (\ref{eq:weight2}).

\subsection{A geometrical diagnostic of dark energy}

Cosmological observations made during the past two decades have brought about 
both qualitative and quantitative changes in our perception of the Universe.
Observational support has come for many theories (such as inflation) which 
earlier were regarded as being at the level of hypothesis. On the other hand, 
fundamentally new properties of our Universe, such its current
state of acceleration, have also been unravelled. The remarkable qualitative
similarity between current acceleration (fueled by DE) and inflationary 
acceleration (supported by what may be called primordial dark energy) may help 
us understand in which direction cosmology may be headed. The discovery by 
COBE of an approximately scale-invariant spectrum of primordial fluctuations 
($n_s-1 \simeq 0$) was regarded by many as providing tacit support for the 
inflationary scenario. However, virtually all models of inflation predict 
small departures from the scale invariance $n_s - 1 = \epsilon$, where 
$\epsilon$ is a model dependent quantity. Strict scale invariance ($n_s = 1$) 
arises for a specific 2-parameter family of inflaton potentials only 
\cite{st05}, which reduces to $V(\phi)\propto \phi^{-2}$ in the slow-roll
approximation. In this context, the recent 3-year data
release of the WMAP experiment suggests $n_s-1 \simeq -0.04$, which is in 
excellent agreement with predictions made by the simplest inflationary models 
such as chaotic inflation with $V(\phi)\propto \phi^2$.

The present situation concerning DE resembles, in some respects, the status of 
inflation just after the release of the COBE data in 1992. The cosmological 
constant is in excellent agreement with data and departures (if any) from 
$1+w = 0$ are believed to be quite small. Nevertheless, since at present no 
fundamental theory predicting the (small) value of the $\Lambda$-term 
exists, it is of utmost importance that departures from $1+w = 0$ 
be probed to accuracies of at least $1\%$ by future generations of \de
experiments. Since virtually all DE models (other than $\Lambda$) have either 
$w_0 \neq -1$ or ${\dot w_0} \neq 0$, the need for probing both the equation 
of state as well as its first derivative become crucial if any deep insight is 
to be gained into the nature of DE.

It is useful to recall that several DE models which agree well with the 
current data arise because of modifications to the gravitational sector of 
the theory \cite{DE_review,DGP,ss02,alam06,maartens06}.
For these {\em geometrical DE} models, the EOS no longer plays the role of a fundamental physical 
quantity
and it would be very useful if we could supplement it with a diagnostic which 
could unambiguously probe the properties of all classes of DE models.
The $w${\tt -probe} discussed earlier, provides one such method, since it is 
based on the expansion history $H=\dot a/a$. Since the expansion factor $a(t)$ 
is an essential feature of all metric theories of gravity, it is worthwhile 
trying to explore the properties of DE by considering the generic form 
\beq\label{eq:taylor0}
a(t) = a(t_0) + {\dot a}\big\vert_0 (t-t_0) +
\frac{\ddot{a}\big\vert_0}{2} (t-t_0)^2 +
\frac{\atridot\big\vert_0}{6} (t-t_0)^3 + ...~.
\eeq
Cosmic acceleration appears to be a fairly recent phenomenon
\cite{benitez,riess01,alam04a,alam04b}, so we can confine our attention to 
small values of $\vert t-t_0\vert$ in (\ref{eq:taylor0}). The second and third 
terms in the RHS of (\ref{eq:taylor0}) have played a crucial
role in the development of cosmology. Indeed, for a long time the large errors
in $H = {\dot a}/a$ and $q = -\ddot{a}/a H^2$ impaired a precise picture of
cosmic expansion and only recently has the sign of $q$ been determined to 
sufficient accuracy for us to make the statement that the universe is 
accelerating.
\footnote{The geometrical relation $R/6H^2 = \Omega_{\rm total} - q$ links the Ricci 
scalar $R$ to $q$. Therefore, an accelerating ($q<0$) spatially flat universe 
corresponds to $R/6H^2>1$.}
Keeping in mind the significant progress expected in observational cosmology 
over the next decade, we feel the time to be ripe to supplement $H$ and $q$ 
with $r = \atridot/a H^3$, which is the next logical step
in the hierarchy of cosmological parameters. It is remarkable that $r=1$ for 
the spatially flat $\Lambda$CDM model which is the `standard' cosmological 
model at present, as well as for the CDM model. Supplementing $r$ (the first 
statefinder) with the second statefinder $s= (r-1)/3(q-1/2)$ permits us to 
break this degeneracy and to characterize different DE models 
in a very informative manner. \footnote{The first statefinder $r$ was also 
dubbed the cosmic jerk in \cite{visser03}.}

\begin{figure*}
\centerline{ \psfig{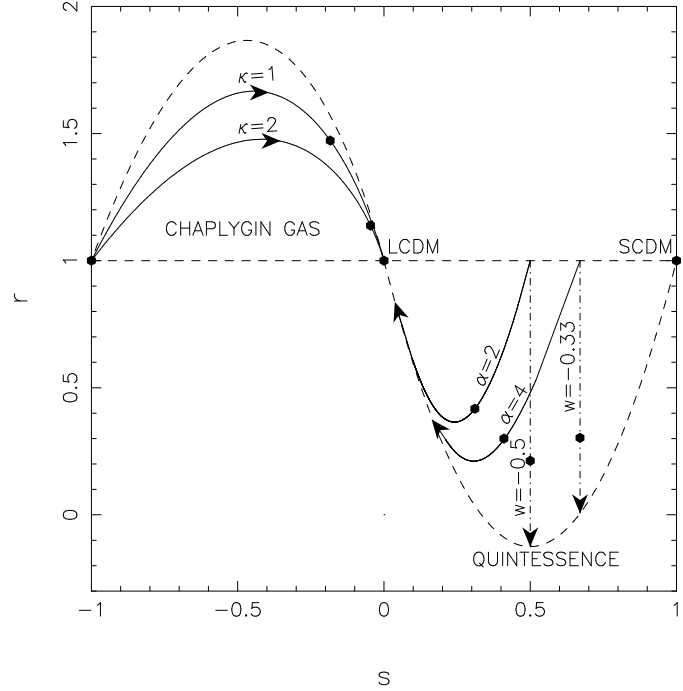} }
\caption{\footnotesize
The time evolution of the statefinder pair
$\lbrace r,s \rbrace$ for quintessence models and the Chaplygin gas.
Solid lines to the right of LCDM represent the inverse power-law potentials
~\protect\cite{ratra88}
$V=V_0/\phi^{\alpha}$, while those to the left correspond to the Chaplygin gas.
Dot-dashed lines represent DE with a
constant equation of state $w$.
Tracker models tend to approach the LCDM fixed
point ($r=1, s=0$) from the right at $t \to \infty$,  whereas the Chaplygin gas
approaches LCDM from the left.
For Chaplygin gas
$\kappa$ is the ratio between matter density and the
density of the Chaplygin gas at early times ~\protect\cite{chap}.
The dashed curve in the lower right is the envelope of all
quintessence models, while the dashed curve in the upper left is the
envelope of Chaplygin gas models (the latter is described by $\kappa =
\Omega_m/1-\Omega_m$).  The region outside the dashed curves is forbidden for
both classes of dark energy models.
The ability of the statefinder to differentiate between dark energy models
is clearly demonstrated.
From Alam \etal ~\protect\cite{statefinder2}.
}
\label{fig:statefinder}
\end{figure*}

The statefinder pair $\statei$ is a `geometrical' diagnostic since it depends 
upon $a(t)$ and hence upon the space-time geometry.
An important property of the statefinder is that spatially
flat LCDM corresponds to the fixed point
\beq
\statei\bigg\vert_{\rm LCDM} = \lbrace 1,0 \rbrace ~.  
\label{eq:fixedpt}
\eeq
So the basic question of whether 
DE is the $\Lambda$-term or `something else' can be rephrased 
into whether the equality (\ref{eq:fixedpt}) is satisfied for DE.
Indeed, the departure of a DE model from 
$\statei = \lbrace 1,0 \rbrace$ provides a good way of
establishing the `distance' of this model from LCDM \cite{statefinder1,statefinder2}. The fact that different classes of DE models show distinctly 
different behaviour when plotted in the $\statei$-plane adds to the practical 
utility of this diagnostic. See figure \ref{fig:statefinder} and 
\cite{statefinder2,statefinder_papers} for some applications of the 
statefinder.

Finally, the pair $\statei$ can also be expressed in terms of 
$\lbrace w, {\dot w}\rbrace$ quite simply
\ber\label{eq:statefinder1}
r &=& 1 + \frac{9w}{2} \omx(1+w) -
\frac{3}{2} \omx \frac{\dot{w}}{H} \,\,,\\
s &=& 1+w-\frac{1}{3} \frac{\dot{w}}{wH}\,\,.
\eer
It is instructive to note that the second statefinder can be rewritten as 
\cite{statefinder2}
\beq
s={(\rho + p)\over p}
{\dot {p} \over \dot{\rho}}~
\label{eq:statefinder2}
\eeq
where $p = \sum_a p_a$ is the {\em total} pressure including that of DE. From 
(\ref{eq:statefinder2}) we find that $s$ is very sensitive to the epoch when
the total pressure in the Universe vanishes. For $\Lambda$CDM, this takes 
place at the redshift $z_p \simeq 10$, when the positive radiation pressure 
cancels the negative pressure in $\Lambda$. Consequently, $s \to \infty$ as 
$z \to z_p$. In general, the redshift $z_p$ at which the total pressure in the 
Universe becomes zero is quite sensitive to the nature of DE.

\begin{table*}[tbh!]
\begin{center}
\begin{minipage}[h]{0.9\linewidth} \mbox{} \vskip -18pt
\bigskip
\begin{tabular}{lll} 
Level & Geometrical Parameter & Physical Parameter \\\hline
 \\
 1         &   $H(z) \equiv \frac{\dot a}{a}$  &
~~$\rho_m(z) = \rho_{0m} (1+z)^3$, \\
& & ~~$\rho_{\rm DE} = \frac{3H^2}{8\pi G} - \rho_m$\\
\\
\hline
\\
2          &   
$q(z) \equiv -\frac{{\ddot a}a}{{\dot a}^2} = -1 + \frac{d\log{H}}
{d\log{(1+z)}}$ &
   ~~$V(z),~~T(z) \equiv \frac{{\dot\phi}^2}{2}~,~~~w(z) = 
\frac{T - V}{T + V}~, $\\
& $q(z)\bigg\vert_{\Lambda{\rm CDM}} = -1 + \frac{3}{2}\Omega_m(z)$ &  
~~$\Omega_V = \frac{8\pi G V}{3H^2}~,~~\Omega_T = \frac{8\pi G T}{3H^2} $\\
\\
\hline
\\
3 & $r(z) \equiv \frac{\atridot a^2}{{\dot a}^3}, ~s \equiv \frac{r-1}
{3(q-1/2)}$ & ~~$\Pi(z) \equiv {\dot V} = {\dot \phi}V'$, ~~$\Omega_\Pi = \frac{8\pi G{\dot V}}{3H^3}$\\
\\
&$\statei\bigg\vert_{\Lambda{\rm CDM}} = \lbrace 1,0 \rbrace $
 & \\
\\
\hline
\end{tabular}
\label{table:dark}
\end{minipage}
\end{center}
\medskip
\end{table*}

Finally, returning to our analogy between an inflaton and dark energy, an
interesting correspondence exists between the $\lbrace{q,r}\rbrace$ pair
describing DE and the {\em slow roll parameters} $\epsilon = - {\dot H}/H^2$,
$\eta = -{\ddot H}/2H{\dot H}$, describing inflation
\ber
q + 1 &=& \epsilon\nonumber\\
r - 1 &=& \epsilon\left (2\eta-3\right ) ~.
\eer

We end this section by noting that our current description of cosmology
relies on parameters which are either {\em geometrical} or {\em physical} in
nature. In the table above, we have divided cosmological parameters according to
their {\em level} which is related to the number of differentiations of the
expansion factor needed to construct that parameter. Note that all relative
energy densities in the table $\Omega_a(z)$ are defined using the present
value $G=G_0$ of the Newton gravitational constant in agreement with the
definition (\ref{Einst}). Each higher level requires an additional
differentiation of observational data $(D_L, D_A,...)$ and, therefore, demands
a higher level of accuracy for the latter. For current data, we broadly have
about 10\% accuracy for quantities belonging to the first level, 50\% for the
second level, while determination of third level parameters lies in the future.
By smoothing, we lower the level of a parameter by unity.

Relations between geometrical and physical parameters are the following:

\ber
w(z) = {2 q(z) - 1 \over 3 \left( 1 - \Omega_{\rm m}(z) \right)}~; \\
\nonumber
\Omega_V(z)  = \frac{2-q}{3} - \frac{H_0^2}{2H^2}\Omega_{m0}(1+z)^3~;\\
\nonumber
\Omega_T(z)  = \frac{1+q}{3} - \frac{H_0^2}{2H^2}\Omega_{mo}(1+z)^3~; \\
\nonumber
\Omega_{\Pi}(z) = \frac{1}{3}\left( r - 3q - 4 + \frac{9H_0^2}{2H^2}\Omega_{m0}
(1+z)^3\right)~.
\label{relation}
\eer

\section{Summary and discussion}

The nature of dark energy is clearly one of the outstanding physical and
cosmological puzzles of this century. Although many distinct theoretical 
models have been advanced to explain the cosmic acceleration, an alternative 
approach to study and understand DE is to determine DE properties from 
observational data in a model independent fashion, thus, {\em reconstructing}
DE from observations. In this review, we have attempted to briefly summarize 
some important approaches of cosmological reconstruction. These approaches are briefly 
classifiable into parametric and non-parametric methods. Both methods appear 
to give consistent results when applied to current data. Future directions for 
model independent reconstruction of dark energy have also been briefly discussed.
In the first approximation, within current observational errors, the 
reconstructed DE behaves like a cosmological constant. But future, much more 
precise data, may well show some deviations from this behaviour. However small, 
these could have a profound influence on the whole of physics.

\bigskip

{\em Acknowledgments\/}: We are grateful to Dragan Huterer, 
Irith Maor, Yun Wang and Jun-Qing Xia for agreeing to provide figures
for this review. VS acknowledges a useful conversation with Ranjeev Misra.
AS was partially supported by the Russian Foundation for Basic Research, 
grant 05-02-17450, and by the Research Program ``Astronomy" of the Russian 
Academy of Sciences. He also thanks the Centre Emile Borel, Institut Henri 
Poincare, Paris for hospitality in the period when this paper was finished.

\end{document}